\documentclass{emulateapj}

\slugcomment{ApJ Letters, in press}

\shorttitle{Heavily Obscured AGN in high-z LIRGs}
\shortauthors{Treister et al.}

\begin{document}


\title{Heavily Obscured AGN in High Redshift Luminous Infrared Galaxies}


\author{Ezequiel Treister\altaffilmark{1,2}, C. Megan Urry\altaffilmark{3,4,5},  Kevin Schawinski\altaffilmark{2,4,5}, Carolin N. Cardamone\altaffilmark{3,4},
David B. Sanders\altaffilmark{1}}

\altaffiltext{1}{Institute for Astronomy, 2680 Woodlawn Drive, University of Hawaii, Honolulu, HI 96822; treister@ifa.hawaii.edu}
\altaffiltext{2}{Chandra/Einstein Fellow}
\altaffiltext{3}{Department of Astronomy, Yale University, PO Box 208101, New Haven, CT 06520.}
\altaffiltext{4}{Yale Center for Astronomy and Astrophysics, P.O. Box 208121, New Haven, CT 06520.}
\altaffiltext{5}{Department of Physics, Yale University, P.O. Box 208121, New Haven, CT 06520.}

\begin{abstract}
We take advantage of the rich multi-wavelength data available in the Chandra Deep Field South (CDF-S), including the 4 Msec Chandra observations (the deepest 
X-ray data to date), in order to search for heavily-obscured low-luminosity AGN among infrared-luminous galaxies. In particular, we obtained a stacked rest-frame
X-ray spectrum for samples of galaxies binned in terms of their IR luminosity or stellar mass. We detect a significant signal at E$\sim$1 to 8 keV, which we interpret as 
originating from a combination of emission associated with star-formation processes at low energies combined with a heavily-obscured AGN at E$>$5~keV. We further find that the relative strength of 
this AGN signal decays with decreasing IR luminosity, indicating a higher AGN fraction for more luminous IR sources. Together, these results strongly
suggest the presence of a large number of obscured AGN in IR-luminous galaxies. Using samples binned in terms of stellar mass in the host
galaxy, we find a significant excess at E=6-7~keV for sources with M$>$10$^{11}$M$_\odot$, consistent with a large obscured AGN population
in high mass galaxies. In contrast, no strong evidence of AGN activity was found for less-massive galaxies. The integrated intensity at high energies indicates that a 
significant fraction of the total black hole growth, $\sim$22\%, occurs in heavily-obscured systems that are not individually detected in even the deepest X-ray 
observations. There are also indications that the number of low-luminosity, heavily-obscured AGN does not evolve significantly with redshift, in contrast to 
the strong evolution seen in higher luminosity sources.
\end{abstract}


\section{Introduction}

Most of the accretion onto the supermassive black hole (SMBH) found in the center of most massive galaxies is heavily 
obscured by the surrounding dust and gas (e.g., \citealp{fabian99a}). In the local Universe, $\sim$75\% of the Seyfert 2 galaxies
are heavily-obscured ($N_H$$>$10$^{23}$~cm$^{-2}$; \citealp{risaliti99}). Many of these, if at $z$$\gtrsim$1, where most of the black hole growth
occurs, would not be identified in X-rays even in very deep ($>$1 Msec) Chandra or XMM/Newton exposures \citep{treister04}. Locating and 
quantifying this heavily obscured SMBH growth, in particular at high redshifts, is currently one of the fundamental problems 
in astrophysics. 

Because the energy absorbed at optical to X-ray wavelengths is later re-emitted in the mid-IR, it is expected that all Active Galactic Nuclei (AGN), even
the most obscured ones, should be very bright mid-IR sources (e.g., \citealp{martinez06}). Hence, it is not surprising
that a large number of heavily obscured --- even Compton-thick ($N_H$$>$10$^{24}$cm$^{-2}$) --- AGN have been found amongst the Luminous and
Ultra-luminous Infrared Galaxies ((U)LIRGs; L$_{IR}$$>$10$^{11}$ and $>$10$^{12}$L$_\odot$ respectively), both locally \citep{iwasawa09} and at 
high redshift \citep{bauer10}. Deep X-ray observations performed using the XMM-Newton (e.g., \citealp{braito03,braito04}), Chandra \citep{teng05} and 
Suzaku \citep{teng09} observatories have shown that most ULIRGs are intrinsically faint X-ray sources, most likely due to the effects of obscuration, while 
their X-ray spectra show combined signatures of starburst and AGN activity. The key features observed in the X-ray spectra of ULIRGs are 
a soft thermal component, typically associated with star formation, a heavily-obscured (N$_H$$\sim$10$^{24}$ cm$^{-2}$) power-law associated 
with the AGN direct emission, and a prominent emission line at $\sim$6.4~keV, identified with fluorescence emission from iron in the 
K$_\alpha$ ionization level, originating either in the accretion disk or in the surrounding material \citep{matt91}. 

The presence of heavily-obscured AGN among the most extreme ULIRGs at $z$$\simeq$1-2 has recently been established from deep
Spitzer observations \citep{daddi07,fiore08,treister09c}. Most of these sources have very high, quasar-like, intrinsic luminosities, and hence most likely
do not constitute the bulk of the heavily-obscured AGN population \citep{treister10}. Establishing the fraction of (U)LIRGs that host a lower luminosity AGN is 
a more challenging task. Recent works based on X-ray stacking \citep{fiore09} and using 70-$\mu$m selected sources \citep{kartaltepe10} report 
a steep decrease in the fraction of AGN with decreasing IR luminosity, going from $\sim$100\% at L$_\textnormal{IR}$=10$^{13}$~L$_\odot$ 
to $<$10\% at L$_\textnormal{IR}$=10$^{10}$~L$_\odot$. In the local Universe, \citet{schawinski10} found that the incidence of low-luminosity, Seyfert-like, 
AGN as a function of stellar mass is more complicated, and is influenced by other parameters. For example, the dependence of AGN fraction on stellar mass
can be opposite if galaxy morphology is considered (increases with decreasing mass in the early-type galaxy population).

In this work, we estimate the fraction of heavily-obscured AGN in mid-IR-luminous and massive galaxies at high redshift, few of which are individually 
detected in  X-rays. The main goal is to constrain the amount of obscured SMBH accretion happening in distant galaxies. This can be done thanks
to the very deep X-ray observations available in the Chandra Deep Fields and the very low and stable Chandra background, which allows for the efficient
stacking of individually undetected sources. Throughout this letter, we assume a $\Lambda$CDM cosmology with $h_0$=0.7, $\Omega_m$=0.27
and $\Omega_\Lambda$=0.73, in agreement with the most recent cosmological observations \citep{hinshaw09}.

\section{Analysis and Results}

By stacking individually-undetected sources selected at longer wavelengths, it is possible to detect very faint X-ray emitters using Chandra observations. For 
example, this technique was used successfully by \citet{brandt01} in the Chandra Deep Field North (CDF-N) to measure
the average X-ray emission from a sample of Lyman break galaxies at $z$$\simeq$2--4 and by \citet{rubin04} to detect
X-rays from red galaxies at $z$$\sim$2. More recently, samples of heavily-obscured AGN candidates selected based
on their mid-IR properties have been studied in X-rays via Chandra stacking (e.g., \citealp{daddi07,fiore08,treister09c}).

The 4 Msec Chandra observations of the Chandra Deep Field South (CDF-S), are currently
the deepest view of the X-ray sky. In addition, the CDF-S has been observed extensively at many wavelengths. The multiwavelength data available on the (E)CDF-S were 
presented by \citet{treister09c}. Very relevant for this work are the deep Spitzer observations available in this field, using both the Infrared Array Camera (IRAC) and the 
Multiband Imaging Photometer for Spitzer (MIPS), from 3.6 to 24~$\mu$m. Also critical is the availability of good quality photometric 
redshifts ($\Delta$$z$/(1+$z$)=0.008 for $R$$<$25.2) obtained thanks to deep observations in 18 medium-band optical filters performed using 
Subaru/Suprime-Cam \citep{cardamone10}. 

We generated our sample starting with the 4959 Spitzer/MIPS 24~$\mu$m sources in the region covered by the Chandra observations that have
photometric redshift $z$$>$0.5, and hence rest-frame E$>$4~keV emission falling in the high-sensitivity Chandra range. In addition, sources 
individually detected in X-rays and reported in the catalogs of \citet{luo08}, \citet{lehmer05} or \citet{virani06} were removed from our 
sample, as these sources will otherwise dominate the stacked spectrum. We then inspected the remaining sources to eliminate
individual detections in the 4 Msec data not present in the 2 Msec catalog of \citet{luo08}. We further excluded 28 sources that meet the 
selection criteria of \citet{fiore08} for heavily obscured AGN, $f_{24}$/$f_R$$>$1000 and $R$-$K$$>$4.5 (Vega), because we expect these sources to 
contain an intrinsically luminous AGN (quasar), while the aim of this work is to find additional hidden accretion in less luminous objects. The median 
redshift of the sources in our final sample is 1.32 (average $z$=1.5) with a standard deviation of 0.77.

In order to perform X-ray stacking in the rest-frame, we started with the regenerated level 2 merged event files created by the Chandra X-ray 
Center\footnote{Data publicly available at http://cxc.harvard.edu/cda/whatsnew.html}. For each source, we extracted all events in a circle of 
30$''$ radius centered in the optical position. The energy of each event was then converted to the rest frame using the photometric redshift of the 
source. Using standard CIAO \citep{fruscione06} tools we then generated seven X-ray images for each source covering the energy range from 1-8 keV 
in the rest-frame with a fixed width of 1 keV. Images for individual sources were then co-added to measure the stacked signal. Total counts were measured
in a fixed 5$''$ aperture, while the background was estimated by randomly placing apertures with same the area, 5$''$ to 30$''$ away from the center.

Several groups have found (e.g, \citealp{kartaltepe10} and references therein) that the fraction of galaxies containing an AGN is a strong function of
their IR luminosity. Hence, it is a natural choice to divide our sample in terms of total IR luminosity. The infrared luminosity was estimated from the 
observed 24~$\mu$m luminosity assuming the relation found by \citet{takeuchi05}: $\log$(L$_{IR}$)=1.02+0.972 $\log$(L$_{12~\mu m}$). We further 
assumed that the $k$ correction between observed-frame 24~$\mu$m and rest-frame 12~$\mu$m luminosity for these sources  
is negligible, as shown by \citet{treister09c}. We then separated our sample in 4 overlapping bins: $L_{IR}$$>$10$^{11}$$L_\odot$, 
$L_{IR}$$>$5$\times$10$^{10}$$L_\odot$, 5$\times$10$^{10}$$L_\odot$$>$$L_{IR}$$>$10$^{10}$$L_\odot$ and $L_{IR}$$>$10$^{10}$$L_\odot$ 
and stacked them independently. The number of sources in each sample is 670, 1545, 2342 and 3887 respectively.

\begin{figure}
\begin{center}
\includegraphics[width=0.22\textwidth]{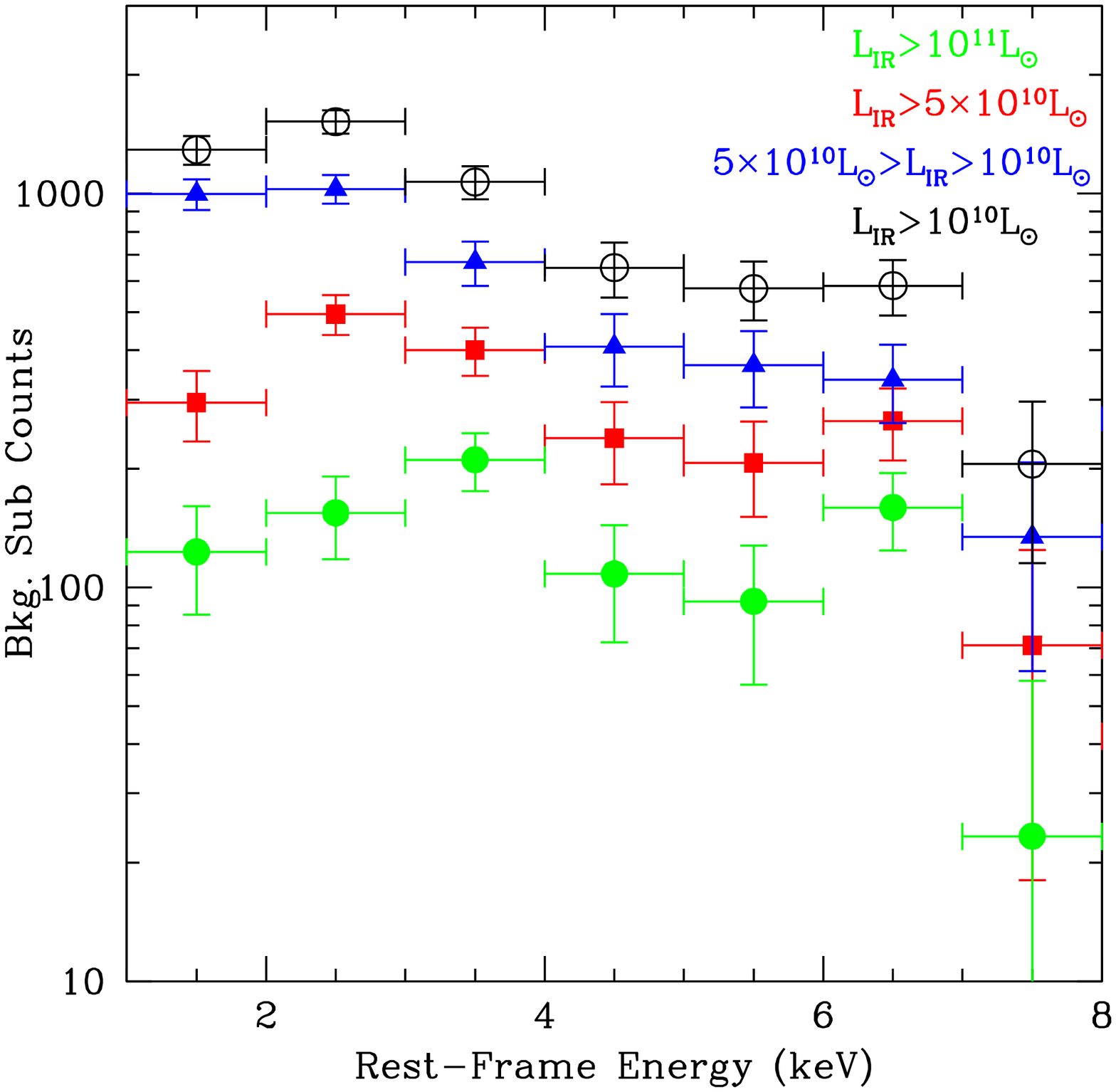}
\includegraphics[width=0.22\textwidth]{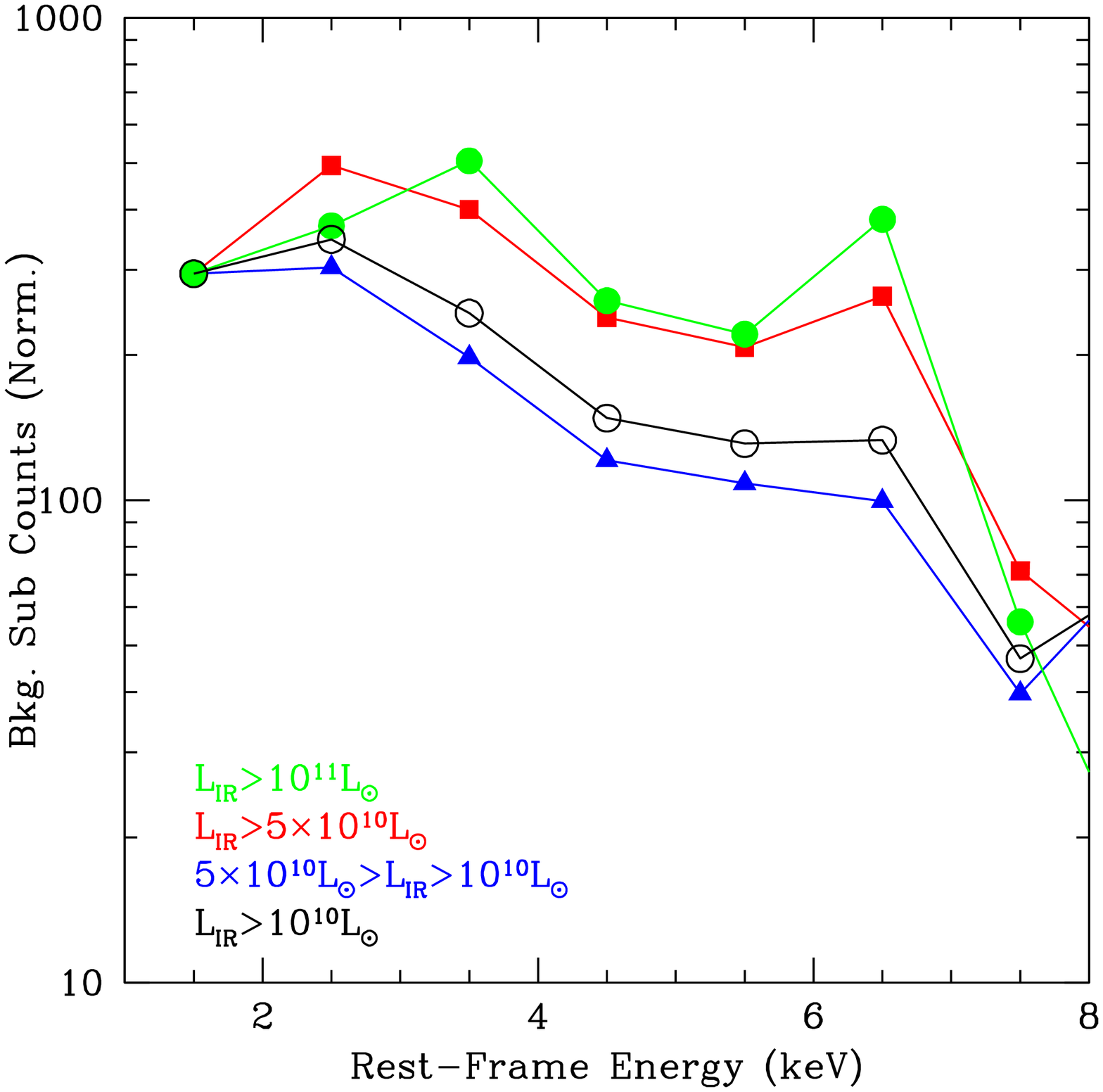}
\end{center}
\caption{{\it Left panel:} Stacked background-subtracted Chandra counts as a function of rest-frame energy from 1 to 8 keV. Samples were selected
based on their IR luminosity in the following overlapping bins: $L_{IR}$$>$10$^{11}$$L_\odot$ ({\it filled circles}), $L_{IR}$$>$5$\times$10$^{10}$$L_\odot$ 
({\it squares}), 5$\times$10$^{10}$$L_\odot$$>$$L_{IR}$$>$10$^{10}$$L_\odot$ ({\it triangles}) and $L_{IR}$$>$10$^{10}$$L_\odot$ ({\it open circles}).
{\it Right panel}: same as left panel but normalized at 1 keV in order to highlight the differences in spectral shape among the different samples.
The largest differences are at E$\gtrsim$5 keV, where there is a clear trend in the relative intensity as a function of IR luminosity,
suggesting a larger fraction of AGN in the most luminous IR sources.}
\label{obs_spec}
\end{figure}

In Fig.~\ref{obs_spec} we present the stacked spectra as a function of rest-frame energy, both in total counts and normalized at 1 keV to highlight the 
difference in spectral shape among the different samples. At E$\gtrsim$5 keV, the spectra begin to diverge, where we expect the AGN emission to
dominate even for heavily-obscured sources. There is a clear trend, with more high energy X-ray emission with increasing IR luminosity.

\section{Discussion}

The spectra shown in Fig.~\ref{obs_spec} cannot be directly interpreted, as the detector-plus-telescope response information is lost after the conversion to 
rest-frame energy and stacking. Hence, we perform simulations assuming different intrinsic X-ray spectra in order to constrain the nature of the sources 
dominating the co-added signal. We use the XSPEC code \citep{arnaud96} to convolve several intrinsic input spectra with the latest response 
functions\footnote{Obtained from http://cxc.harvard.edu/caldb/calibration/acis.html} for the Chandra ACIS-I camera used in the CDF-S observations. We
then compare these simulated spectra with the observations in our sample of IR-selected sources.

The low energy spectrum of (U)LIRGs is dominated by a combination of a thermal plasma component with temperatures $kT$$\simeq$0.7~keV, particularly
important at E$<$3 keV, and the emission from high-mass X-ray binaries (HMXBs) at 1$<$E (keV)$<$10 (e.g., \citealp{persic02}). For each source, we
generated a simulated spectrum using a combination of these two components, taking into account the luminosity and redshift of the source. For
the HMXB population we assumed a power-law given by $\Gamma$=1.2 and cutoff energy $E_c$=20 keV, consistent with recent observations 
(e.g., \citealp{lutovinov05}). This component was normalized assuming the relation between IR and X-ray luminosity in starburst galaxies found 
by \citet{ranalli03}. For the thermal component, we assumed a black body with temperature $kT$=0.7 keV. The normalization of this component
was then adjusted to match the observations at $E$$<$3 keV.

\begin{figure}
\begin{center}
\plotone{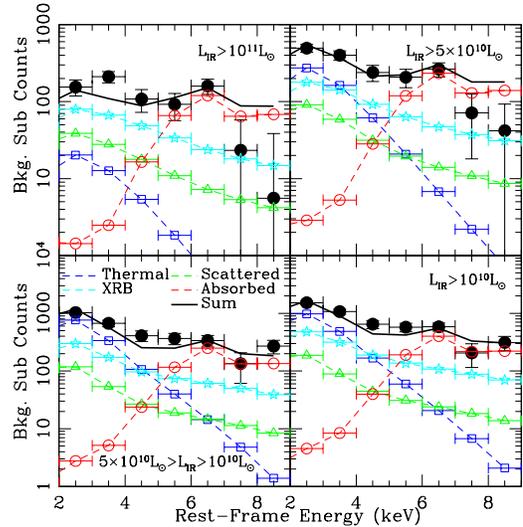}
\end{center}
\caption{Stacked background-subtracted Chandra counts as a function of rest-frame energy, as in Fig.~\ref{obs_spec}. {\it Black data points (filled circles)} show
the stacked X-ray signal for sources binned by IR luminosity. The {\it cyan dashed lines (stars)} shows the simulated spectra for the HMXB population
normalized using the \citet{ranalli03} relation between star-formation rate and X-ray luminosity. The {\it blue dashed lines (open squares)} show simulated thermal spectra
corresponding to a black body with $kT$=0.7 keV, required to explain the E$<$3 keV emission. An absorbed AGN spectrum, given by a power-law with $\Gamma$=1.9 and a 
fixed $N_H$=10$^{24}$~cm$^{-2}$, is shown by the {\it red dashed lines (open circles)}. In addition, a scattered AGN component, characterized by a 1\% reflection of the underlying 
unobscured power-law, is shown by the {\it green dashed lines (open triangles)}. The resulting summed spectrum ({\it black solid lines}) is in very good 
agreement with the observed counts. The strong detection in the stacked spectrum at E$>$5 keV, in particular at the higher IR luminosities, confirms the presence of a significant 
number of heavily-obscured AGN in these samples.}
\label{simul_spec}
\end{figure}

In order to compute the possible contribution from heavily-obscured AGN to the stacked spectrum we assumed the observed X-ray spectrum of the nearby
ULIRG IRAS19254-7245, as observed by Suzaku \citep{braito09}. In addition to the starburst emission described above, the X-ray spectrum is described by
an absorbed, Compton-thick, power-law with $\Gamma$=1.9, $N_H$=10$^{24}$~cm$^{-2}$, and a possible scattered component, characterized by a 
power-law with $\Gamma$=1.9, no absorption, and 1\% of the intensity of the direct emission. The resulting simulated spectral components and the comparison
with the observed stacked spectrum for sources with the four samples defined above are shown in Fig.~\ref{simul_spec}. 

It is not possible to explain the observed stacked spectral shape using only a plausible starburst spectrum without invoking an AGN component, which dominates 
at E$>$5~keV. The average intrinsic rest-frame 2-10 keV AGN luminosity needed to explain the observed spectrum, assuming that every source in the sample contains an AGN of the 
same luminosity, is 6$\times$10$^{42}$~erg~s$^{-1}$ for sources with $L_{IR}$$>$10$^{11}$$L_\odot$, 3$\times$10$^{42}$~erg~s$^{-1}$ 
for sources with $L_{IR}$$>$5$\times$10$^{10}$$L_\odot$, 5$\times$10$^{41}$~erg~s$^{-1}$ in the sample with 5$\times$10$^{10}$$L_\odot$$>$$L_{IR}$$>$10$^{10}$$L_\odot$
and 7$\times$10$^{41}$~erg~s$^{-1}$ for sources with $L_{IR}$$>$10$^{10}$$L_\odot$. All of these are (intrinsically) very low-luminosity AGN; even if there is a range, it is extremely
unlikely to include high-luminosity quasars like those discussed in previous stacking papers. An alternative possibility is that the extra emission at E$>$5~keV is due entirely to 
the Fe~K$\alpha$ line, provided the errors in the photometric redshifts in these samples are significantly larger than the values reported by \citet{cardamone10}. 
Regardless of the template assumed for the AGN emission, we obtain similar values for the average AGN luminosity in each sample.

The median hard X-ray luminosity for the Chandra sources with measured photometric redshifts in the catalog of \citet{luo08} is 4.1$\times$10$^{43}$ erg~s$^{-1}$ for 
the sources in the $L_{IR}$$>$10$^{11}$$L_\odot$ sample, 3.5$\times$10$^{43}$~erg~s$^{-1}$ in the $L_{IR}$$>$5$\times$10$^{10}$$L_\odot$ group, 5.7$\times$10$^{42}$~erg~s$^{-1}$ for 
sources with 5$\times$10$^{10}$$L_\odot$$>$$L_{IR}$$>$10$^{10}$$L_\odot$ and 1.6$\times$10$^{43}$~erg~s$^{-1}$ in the $L_{IR}$$>$10$^{10}$$L_\odot$ sample. Hence, 
if the heavily-obscured AGN in our stacked samples have the same median intrinsic luminosity this would indicate that 15\% (98 sources) of the 670 galaxies 
with $L_{IR}$$>$10$^{11}$$L_\odot$ contain a heavily-obscured AGN. This fraction is $\sim$9\% (132 and 205 sources respectively) in the $L_{IR}$$>$5$\times$10$^{10}$$L_\odot$ and
5$\times$10$^{10}$$L_\odot$$>$$L_{IR}$$>$10$^{10}$$L_\odot$ samples. For sources with $L_{IR}$$>$10$^{10}$$L_\odot$ this fraction is $<$5\%. The integrated intrinsic 
X-ray emission in the rest-frame 2-10 keV band due to the heavily-obscured AGN in this sample, obtained by multiplying the intrinsic X-ray luminosity by the number of sources 
and dividing by the studied area, is $\sim$4.6$\times$10$^{46}$~erg~cm$^{-2}$~s$^{-1}$~deg$^{-2}$. For comparison, the total emission from all the X-ray 
detected AGN in the CDF-S is 1.63$\times$10$^{47}$~erg~cm$^{-2}$~s$^{-1}$~deg$^{-2}$. Hence, this extra AGN activity can account for $\sim$22\% of the total SMBH accretion.
Adding this to the obscured SMBH growth in X-ray detected AGN \citep{luo08}, we confirm that most SMBH growth, $\sim$70\%, is significantly obscured and missed by even the 
deepest X-ray surveys \citep{treister04,treister10}.

Performing a similar study on the 28 sources with $f_{24}$/$f_R$$>$1000 and $R$-$K$$>$4.5 that we previously excluded, we find a very hard X-ray
spectrum, harder than that of the $L_{IR}$$>$10$^{11}$$L_\odot$ sources. This spectrum is consistent with a population of luminous AGN with intrinsic rest-frame 2-10 keV 
luminosity $\sim$2$\times$10$^{43}$~erg~s$^{-1}$ and negligible contribution from the host galaxy, except at E$<$2 keV where the thermal component is $\sim$30\% of the
total emission. This result justifies our choice of removing these sources from our study (otherwise they would dominate the stacked signal), while at the same time it confirms the
AGN nature of the vast majority of these sources, in contrast to the suggestion that the extra IR emission could be due to star-formation processes \citep{donley08, pope08,georgakakis10}. 
A similar result for these high-luminosity sources was found by \citet{fiore10}: In a sample of 99 mid-IR excess sources in the COSMOS field he found a strong stacked signal 
at E$\sim$6~keV, which he interpreted as due to the Fe~K$\alpha$ line, a clear signature of AGN emission and high obscuration (see discussion below).

\subsection{Multiwavelength Properties}

By design, none of the sources in our sample are individually detected in X-rays, nor do they satisfy the selection criteria of \citet{fiore08}. However, it is interesting to investigate
if they present other AGN signatures. For example, 237 out of the 1545 sources with $L_{IR}$$>$5$\times$10$^{10}$$L_\odot$ in our sample (15\%) are found inside the
AGN IRAC color-color region defined by \citet{stern05}. For comparison, in the sample of 2342 sources with 5$\times$10$^{10}$$L_\odot$$>$$L_{IR}$$>$10$^{10}$$L_\odot$ --- 
in which from the stacked hard X-ray signal we determined a negligible AGN fraction --- there are 327 galaxies (14\%) in the \citet{stern05} region. This suggests that the 
IRAC color-color diagram cannot be used to identify heavily-obscured low-luminosity AGN, because the near-IR emission in these sources is dominated by the host 
galaxy \citep{cardamone08}. At longer wavelengths, 83 of the 1545 sources with $L_{IR}$$>$5$\times$10$^{10}$$L_\odot$ were detected in the deep VLA observations of the 
CDF-S \citep{kellermann08}. In contrast, only 33 sources in the 5$\times$10$^{10}$$L_\odot$$>$$L_{IR}$$>$10$^{10}$$L_\odot$ sample were detected in these 
observations. Using the $q_{24}$ ratio between 1.4 GHz and 24~$\mu$m flux densities (e.g., \citealp{appleton04}), we find that in the $L_{IR}$$>$5$\times$10$^{10}$$L_\odot$ 
sample, only 14 sources have $q_{24}$$<$-0.23 and can be considered ``radio-loud'' \citep{ibar08}, and in the 5$\times$10$^{10}$$L_\odot$$>$$L_{IR}$$>$10$^{10}$$L_\odot$ 
sample only 10 sources have $q_{24}$$<$-0.23. Hence, we conclude that the fraction of bona fide radio-loud sources is negligible and that in most cases the radio emission 
is produced by star-formation processes.

\subsection{AGN Fraction Versus Stellar Mass}

\begin{figure}
\begin{center}
\includegraphics[scale=0.2]{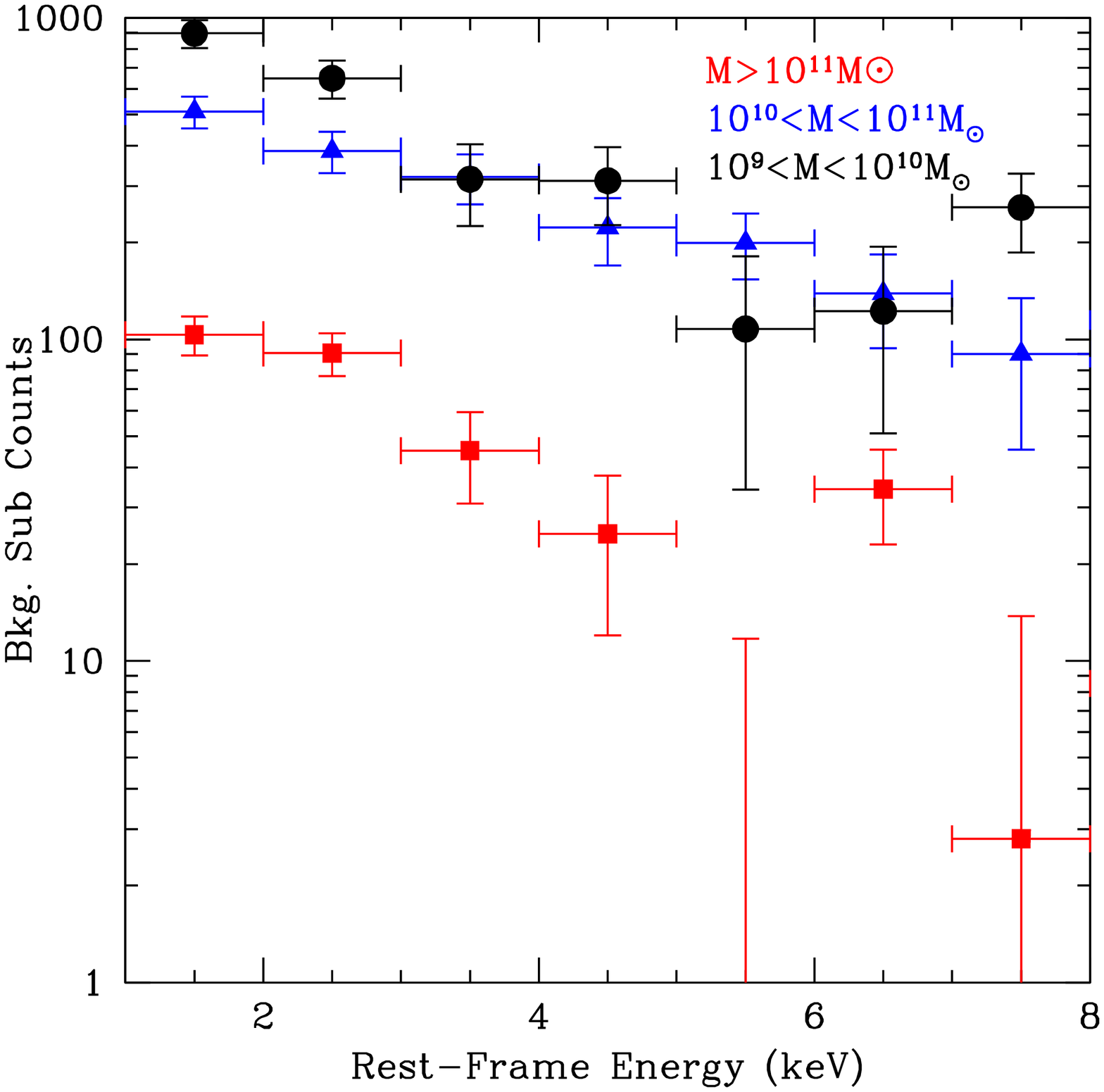}
\includegraphics[scale=0.2]{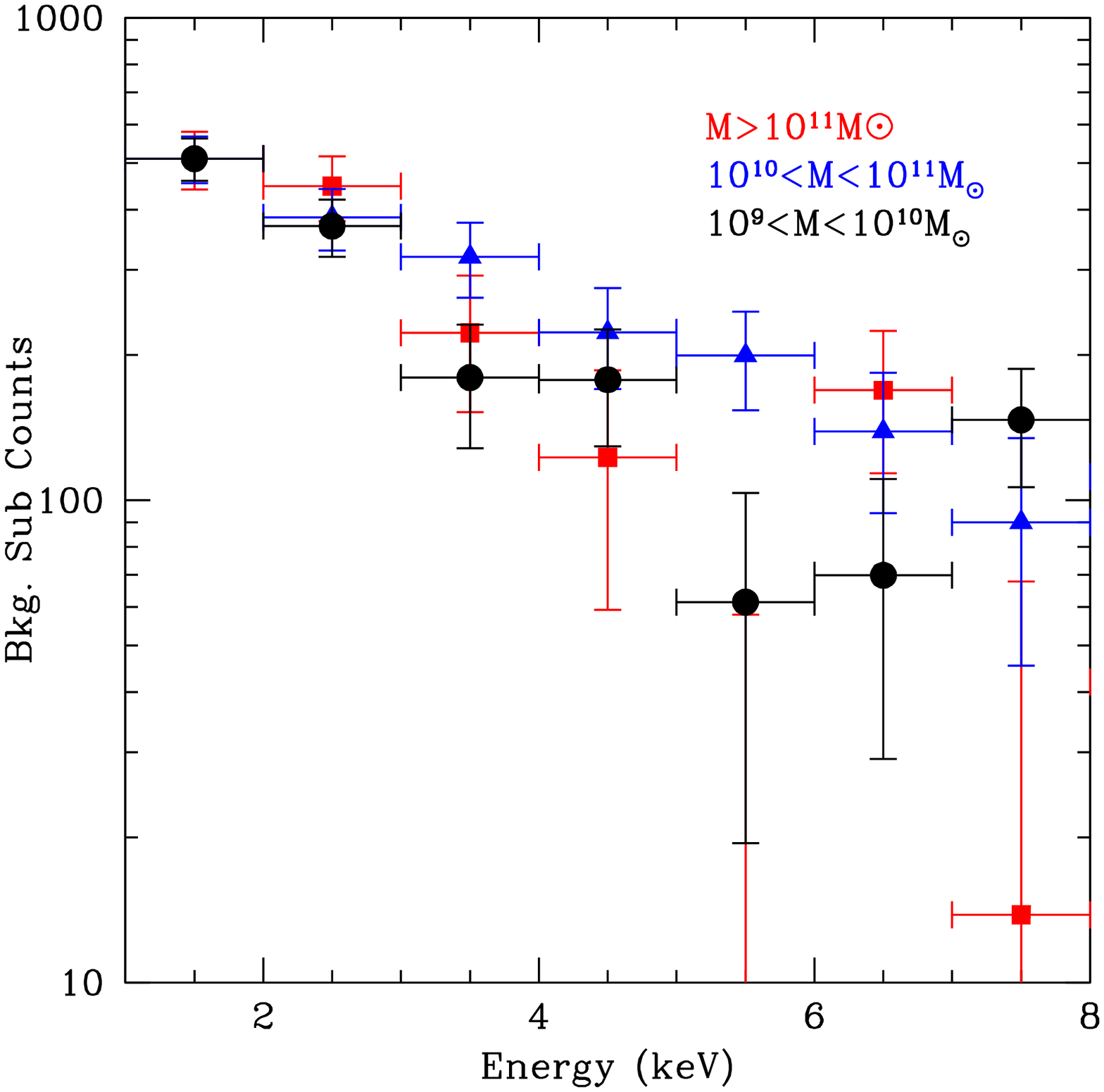}
\end{center}
\caption{Stacked Chandra counts for galaxies binned as a function of their stellar mass. The {\it left panel} shows the spectra for the following
bins: M$>$10$^{11}$M$_\odot$ ({\it red squares}), 10$^{10}$$<$M$<$10$^{11}$M$_\odot$ ({\it blue triangles}) and 10$^{9}$$<$M$<$10$^{10}$M$_\odot$ ({\it black circles}).
{\it Right panel:} same but normalized at 1 keV. In the M$>$10$^{11}$M$_\odot$ sample, the strong excess at E=6-7~keV, which we associate with the Fe~K$\alpha$ line, is an 
indicator of AGN activity. Similarly, for the sources with 10$^{10}$$<$M$<$10$^{11}$M$_\odot$ there is a hard X-ray spectrum, also suggesting a significant AGN fraction.
These preliminary results indicate that these heavily-obscured moderate-luminosity AGN are predominantly present in the most massive galaxies.}
\label{obs_spec_mass}
\end{figure}

In order to investigate the fraction of heavily-obscured AGN as a function of other galaxy parameters, we performed X-ray stacking of samples sorted by stellar mass.
Stellar masses were taken from \citet{cardamone10}, who performed spectral fitting to the extensive optical and near-IR spectro-photometry using FAST \citep{kriek09} and the stellar 
templates of \citet{maraston05} assuming the \citet{kroupa01} initial mass function and solar metallicity. We further restricted our sample to sources with $z$$<$1.2, for 
which photometric redshifts and stellar masses are very well determined ($\Delta$z/(1+$z$)=0.007). We then divided the sample into three mass bins: M$>$10$^{11}$M$_\odot$, 
10$^{11}$$>$M (M$_\odot$)$>$10$^{10}$ and 10$^{10}$$>$M (M$_\odot$)$>$10$^{9}$. The resulting stacked X-ray spectra are shown in Fig.~\ref{obs_spec_mass}.

For sources with M$>$10$^{11}$M$_\odot$, there is a significant excess at 6-7~keV, above a spectrum that otherwise declines with increasing energy. This might 
be due to the presence of the Fe~K$\alpha$ line, a clear indicator of AGN activity.  Contrary to the case of stacking as a function of IR luminosity (Fig.~\ref{simul_spec}), 
here we do not find evidence for an absorbed power-law ---the 6-7 keV feature is simply too sharply peaked. Possibly the restriction to $z$$<$1.2 for the mass-binned stacking, where 
photometric redshifts are most accurate, reveals an emission line that is broadened by less accurate photometric redshifts in the full sample. That is, the feature in the $L_{IR}$-binned stack 
that we interpreted as a heavily absorbed power law may instead be an Fe~K$\alpha$ line broadened artificially by bad photometric redshifts. In the 10$^{11}$$>$M (M$_\odot$)$>$10$^{10}$ 
sample we found a significant hardening of the X-ray spectrum (Fig.~\ref{obs_spec_mass}), suggesting the presence of a significant fraction of AGN. In contrast, only a soft spectrum, 
consistent with star-formation emission, can be seen for sources with 10$^{10}$$>$M (M$_\odot$)$>$10$^{9}$. Taken together, these results indicate that AGN are predominantly present 
in the most massive galaxies, in agreement with the conclusions of \citet{cardamone10b} and others. This will be elaborated in a paper currently in preparation.

\subsection{Space Density of Heavily-Obscured AGN}

The fraction of Compton-thick AGN in the local Universe is still heavily debated. \citet{treister09b} reported a fraction of $\sim$8\% in a flux-limited sample of sources
detected in the Swift/BAT all-sky \citep{tueller08} and  International Gamma-Ray Astrophysics Laboratory (INTEGRAL; \citealp{krivonos07}) surveys. From an INTEGRAL volume-limited 
survey at $z$$<$0.015, \citet{malizia09} found a higher fraction of 24\%, suggesting that even surveys at E$>$10~keV are potentially 
biased against the detection of Compton-thick AGN. The fraction of moderate-luminosity Compton-thick sources in our sample of sources 
with $L_{IR}$$>$5$\times$10$^{10}$$L_\odot$, relative to all AGN in the CDF-S, is $\sim$25\% (132/525), assuming that Compton-thick and Compton-thin AGN have similar 
median  intrinsic luminosities. This indicates that there is no major evolution in the number of moderate-luminosity heavily-obscured AGN from $z$=0 to 2. In contrast, at higher 
luminosities, \citet{treister10} reported that the ratio of obscured to unobscured quasars increased from $\sim$1 at $z$=0 to $\sim$2-3 at $z$$\simeq$2. Hence, although all these 
estimates are still uncertain, it appears that the evolution of Compton-thick AGN depends strongly on their luminosity. We further speculate that this is indication that the 
triggering of low-luminosity AGN is not related to the major merger of gas-rich galaxies as found by \citet{treister10} for high-luminosity quasars or that the time delay
between galaxy interactions and black hole growth is long \citep{schawinski10b}.

\acknowledgements

We thank the referee, Fabrizio Fiore, for very useful and constructive comments. Support for the work of ET and KS was provided by the National
Aeronautics and Space Administration through Chandra/Einstein
Post-doctoral Fellowship Award Numbers PF8-90055 and PF9-00069
respectively issued by the Chandra X-ray Observatory Center, which is
operated by the Smithsonian Astrophysical Observatory for and on
behalf of the National Aeronautics Space Administration under contract
NAS8-03060. CMU and CC acknowledge support from NSF grants 
AST-0407295, AST-0449678, AST-0807570 and Yale University.



\begin{thebibliography}{50}
\expandafter\ifx\csname natexlab\endcsname\relax\def\natexlab#1{#1}\fi

\bibitem[{{Appleton} {et~al.}(2004)}]{appleton04}
{Appleton}, P.~N. {et~al.} 2004, \apjs, 154, 147

\bibitem[{{Arnaud}(1996)}]{arnaud96}
{Arnaud}, K.~A. 1996, in ASP Conf. Ser. 101: Astronomical Data Analysis
  Software and Systems V, ed. G.~H. {Jacoby} \& J.~{Barnes}, 17--+

\bibitem[{{Bauer} {et~al.}(2010){Bauer}, {Yan}, {Sajina}, \&
  {Alexander}}]{bauer10}
{Bauer}, F.~E., {Yan}, L., {Sajina}, A., \& {Alexander}, D.~M. 2010, \apj, 710,
  212

\bibitem[{{Braito} {et~al.}(2004){Braito}, {Della Ceca}, {Piconcelli},
  {Severgnini}, {Bassani}, {Cappi}, {Franceschini}, {Iwasawa}, {Malaguti},
  {Marziani}, {Palumbo}, {Persic}, {Risaliti}, \& {Salvati}}]{braito04}
{Braito}, V., {Della Ceca}, R., {Piconcelli}, E., {Severgnini}, P., {Bassani},
  L., {Cappi}, M., {Franceschini}, A., {Iwasawa}, K., {Malaguti}, G.,
  {Marziani}, P., {Palumbo}, G.~G.~C., {Persic}, M., {Risaliti}, G., \&
  {Salvati}, M. 2004, \aap, 420, 79

\bibitem[{{Braito} {et~al.}(2003){Braito}, {Franceschini}, {Della Ceca},
  {Severgnini}, {Bassani}, {Cappi}, {Malaguti}, {Palumbo}, {Persic},
  {Risaliti}, \& {Salvati}}]{braito03}
{Braito}, V., {Franceschini}, A., {Della Ceca}, R., {Severgnini}, P.,
  {Bassani}, L., {Cappi}, M., {Malaguti}, G., {Palumbo}, G.~G.~C., {Persic},
  M., {Risaliti}, G., \& {Salvati}, M. 2003, \aap, 398, 107

\bibitem[{{Braito} {et~al.}(2009){Braito}, {Reeves}, {Della Ceca}, {Ptak},
  {Risaliti}, \& {Yaqoob}}]{braito09}
{Braito}, V., {Reeves}, J.~N., {Della Ceca}, R., {Ptak}, A., {Risaliti}, G., \&
  {Yaqoob}, T. 2009, \aap, 504, 53

\bibitem[{{Brandt} {et~al.}(2001){Brandt}, {Alexander}, {Hornschemeier},
  {Garmire}, {Schneider}, {Barger}, {Bauer}, {Broos}, {Cowie}, {Townsley},
  {Burrows}, {Chartas}, {Feigelson}, {Griffiths}, {Nousek}, \&
  {Sargent}}]{brandt01}
{Brandt}, W.~N., {Alexander}, D.~M., {Hornschemeier}, A.~E., {Garmire}, G.~P.,
  {Schneider}, D.~P., {Barger}, A.~J., {Bauer}, F.~E., {Broos}, P.~S., {Cowie},
  L.~L., {Townsley}, L.~K., {Burrows}, D.~N., {Chartas}, G., {Feigelson},
  E.~D., {Griffiths}, R.~E., {Nousek}, J.~A., \& {Sargent}, W.~L.~W. 2001, \aj,
  122, 2810

\bibitem[{{Cardamone} {et~al.}(2008){Cardamone}, {Urry}, {Damen}, {van Dokkum},
  {Treister}, {Labb{\'e}}, {Virani}, {Lira}, \& {Gawiser}}]{cardamone08}
{Cardamone}, C.~N., {Urry}, C.~M., {Damen}, M., {van Dokkum}, P., {Treister},
  E., {Labb{\'e}}, I., {Virani}, S.~N., {Lira}, P., \& {Gawiser}, E. 2008,
  \apj, 680, 130

\bibitem[{{Cardamone} {et~al.}(2010a)}]{cardamone10}
{Cardamone}, C.~N. {et~al.} 2010a, \apjs, 189, 270

\bibitem[Cardamone et al.(2010b)]{cardamone10b} Cardamone, C.~N., 
Urry, C.~M., Schawinski, K., Treister, E., Brammer, G., 
\& Gawiser, E.\ 2010b, \apjl~in press, arXiv:1008.2971

\bibitem[{{Daddi} {et~al.}(2007)}]{daddi07}
{Daddi}, E. {et~al.} 2007, \apj, 670, 173

\bibitem[{{Donley} {et~al.}(2008){Donley}, {Rieke}, {P{\'e}rez-Gonz{\'a}lez},
  \& {Barro}}]{donley08}
{Donley}, J.~L., {Rieke}, G.~H., {P{\'e}rez-Gonz{\'a}lez}, P.~G., \& {Barro},
  G. 2008, \apj, 687, 111

\bibitem[{{Fabian} \& {Iwasawa}(1999)}]{fabian99a}
{Fabian}, A.~C. \& {Iwasawa}, K. 1999, \mnras, 303, L34

\bibitem[{{Fiore}(2010)}]{fiore10}
{Fiore}, F. 2010, in American Institute of Physics Conference Series, Vol.
  1248, American Institute of Physics Conference Series, ed. {A.~Comastri,
  L.~Angelini, \& M.~Cappi}, 373--380

\bibitem[{{Fiore} {et~al.}(2008)}]{fiore08}
{Fiore}, F. {et~al.} 2008, \apj, 672, 94

\bibitem[{{Fiore} {et~al.}(2009)}]{fiore09}
---. 2009, \apj, 693, 447

\bibitem[{{Fruscione} {et~al.}(2006)}]{fruscione06}
{Fruscione}, A. {et~al.} 2006, in Presented at the Society of Photo-Optical
  Instrumentation Engineers (SPIE) Conference, Vol. 6270, Society of
  Photo-Optical Instrumentation Engineers (SPIE) Conference Series

\bibitem[{{Georgakakis} {et~al.}(2010){Georgakakis}, {Rowan-Robinson},
  {Nandra}, {Digby-North}, {P{\'e}rez-Gonz{\'a}lez}, \&
  {Barro}}]{georgakakis10}
{Georgakakis}, A., {Rowan-Robinson}, M., {Nandra}, K., {Digby-North}, J.,
  {P{\'e}rez-Gonz{\'a}lez}, P.~G., \& {Barro}, G. 2010, \mnras, 406, 420

\bibitem[{{Hinshaw} {et~al.}(2009)}]{hinshaw09}
{Hinshaw}, G. {et~al.} 2009, \apjs, 180, 225

\bibitem[{{Ibar} {et~al.}(2008){Ibar}, {Cirasuolo}, {Ivison}, {Best}, {Smail},
  {Biggs}, {Simpson}, {Dunlop}, {Almaini}, {McLure}, {Foucaud}, \&
  {Rawlings}}]{ibar08}
{Ibar}, E., {Cirasuolo}, M., {Ivison}, R., {Best}, P., {Smail}, I., {Biggs},
  A., {Simpson}, C., {Dunlop}, J., {Almaini}, O., {McLure}, R., {Foucaud}, S.,
  \& {Rawlings}, S. 2008, \mnras, 386, 953

\bibitem[{{Iwasawa} {et~al.}(2009){Iwasawa}, {Sanders}, {Evans}, {Mazzarella},
  {Armus}, \& {Surace}}]{iwasawa09}
{Iwasawa}, K., {Sanders}, D.~B., {Evans}, A.~S., {Mazzarella}, J.~M., {Armus},
  L., \& {Surace}, J.~A. 2009, \apjl, 695, L103

\bibitem[{{Kartaltepe} {et~al.}(2010)}]{kartaltepe10}
{Kartaltepe}, J.~S. {et~al.} 2010, \apj, 709, 572

\bibitem[{{Kellermann} {et~al.}(2008){Kellermann}, {Fomalont}, {Mainieri},
  {Padovani}, {Rosati}, {Shaver}, {Tozzi}, \& {Miller}}]{kellermann08}
{Kellermann}, K.~I., {Fomalont}, E.~B., {Mainieri}, V., {Padovani}, P.,
  {Rosati}, P., {Shaver}, P., {Tozzi}, P., \& {Miller}, N. 2008, \apjs, 179, 71

\bibitem[{{Kriek} {et~al.}(2009){Kriek}, {van Dokkum}, {Labb{\'e}}, {Franx},
  {Illingworth}, {Marchesini}, \& {Quadri}}]{kriek09}
{Kriek}, M., {van Dokkum}, P.~G., {Labb{\'e}}, I., {Franx}, M., {Illingworth},
  G.~D., {Marchesini}, D., \& {Quadri}, R.~F. 2009, \apj, 700, 221

\bibitem[{{Krivonos} {et~al.}(2007){Krivonos}, {Revnivtsev}, {Lutovinov},
  {Sazonov}, {Churazov}, \& {Sunyaev}}]{krivonos07}
{Krivonos}, R., {Revnivtsev}, M., {Lutovinov}, A., {Sazonov}, S., {Churazov},
  E., \& {Sunyaev}, R. 2007, \aap, 475, 775

\bibitem[{{Kroupa}(2001)}]{kroupa01}
{Kroupa}, P. 2001, \mnras, 322, 231

\bibitem[{{Lehmer} {et~al.}(2005)}]{lehmer05}
{Lehmer}, B.~D. {et~al.} 2005, \apjs, 161, 21

\bibitem[{{Luo} {et~al.}(2008)}]{luo08}
{Luo}, B. {et~al.} 2008, \apjs, 179, 19

\bibitem[{{Lutovinov} {et~al.}(2005){Lutovinov}, {Revnivtsev}, {Gilfanov},
  {Shtykovskiy}, {Molkov}, \& {Sunyaev}}]{lutovinov05}
{Lutovinov}, A., {Revnivtsev}, M., {Gilfanov}, M., {Shtykovskiy}, P., {Molkov},
  S., \& {Sunyaev}, R. 2005, \aap, 444, 821

\bibitem[{{Malizia} {et~al.}(2009){Malizia}, {Stephen}, {Bassani}, {Bird},
  {Panessa}, \& {Ubertini}}]{malizia09}
{Malizia}, A., {Stephen}, J.~B., {Bassani}, L., {Bird}, A.~J., {Panessa}, F.,
  \& {Ubertini}, P. 2009, \mnras, 399, 944

\bibitem[{{Maraston}(2005)}]{maraston05}
{Maraston}, C. 2005, \mnras, 362, 799

\bibitem[{{Mart{\'{\i}}nez-Sansigre} {et~al.}(2006){Mart{\'{\i}}nez-Sansigre},
  {Rawlings}, {Lacy}, {Fadda}, {Jarvis}, {Marleau}, {Simpson}, \&
  {Willott}}]{martinez06}
{Mart{\'{\i}}nez-Sansigre}, A., {Rawlings}, S., {Lacy}, M., {Fadda}, D.,
  {Jarvis}, M.~J., {Marleau}, F.~R., {Simpson}, C., \& {Willott}, C.~J. 2006,
  \mnras, 370, 1479

\bibitem[{{Matt} {et~al.}(1991){Matt}, {Perola}, \& {Piro}}]{matt91}
{Matt}, G., {Perola}, G.~C., \& {Piro}, L. 1991, \aap, 247, 25

\bibitem[{{Persic} \& {Rephaeli}(2002)}]{persic02}
{Persic}, M. \& {Rephaeli}, Y. 2002, \aap, 382, 843

\bibitem[{{Pope} {et~al.}(2008)}]{pope08}
{Pope}, A. {et~al.} 2008, \apj, 689, 127

\bibitem[{{Ranalli} {et~al.}(2003){Ranalli}, {Comastri}, \&
  {Setti}}]{ranalli03}
{Ranalli}, P., {Comastri}, A., \& {Setti}, G. 2003, \aap, 399, 39

\bibitem[{{Risaliti} {et~al.}(1999){Risaliti}, {Maiolino}, \&
  {Salvati}}]{risaliti99}
{Risaliti}, G., {Maiolino}, R., \& {Salvati}, M. 1999, \apj, 522, 157

\bibitem[{{Rubin} {et~al.}(2004){Rubin}, {van Dokkum}, {Coppi}, {Johnson},
  {F{\"o}rster Schreiber}, {Franx}, \& {van der Werf}}]{rubin04}
{Rubin}, K.~H.~R., {van Dokkum}, P.~G., {Coppi}, P., {Johnson}, O.,
  {F{\"o}rster Schreiber}, N.~M., {Franx}, M., \& {van der Werf}, P. 2004,
  \apjl, 613, L5

\bibitem[{{Schawinski} {et~al.}(2010a)}]{schawinski10}
{Schawinski}, K. {et~al.} 2010a, \apj, 711, 284

\bibitem[Schawinski et al.(2010b)]{schawinski10b} Schawinski, K., 
Dowlin, N., Thomas, D., Urry, C.~M., 
\& Edmondson, E.\ 2010b, \apjl, 714, L108 

\bibitem[{{Stern} {et~al.}(2005){Stern}, {Eisenhardt}, {Gorjian}, {Kochanek},
  {Caldwell}, {Eisenstein}, {Brodwin}, {Brown}, {Cool}, {Dey}, {Green},
  {Jannuzi}, {Murray}, {Pahre}, \& {Willner}}]{stern05}
{Stern}, D., {Eisenhardt}, P., {Gorjian}, V., {Kochanek}, C.~S., {Caldwell},
  N., {Eisenstein}, D., {Brodwin}, M., {Brown}, M.~J.~I., {Cool}, R., {Dey},
  A., {Green}, P., {Jannuzi}, B.~T., {Murray}, S.~S., {Pahre}, M.~A., \&
  {Willner}, S.~P. 2005, \apj, 631, 163

\bibitem[{{Takeuchi} {et~al.}(2005){Takeuchi}, {Buat}, {Iglesias-P{\'a}ramo},
  {Boselli}, \& {Burgarella}}]{takeuchi05}
{Takeuchi}, T.~T., {Buat}, V., {Iglesias-P{\'a}ramo}, J., {Boselli}, A., \&
  {Burgarella}, D. 2005, \aap, 432, 423

\bibitem[{{Teng} {et~al.}(2009){Teng}, {Veilleux}, {Anabuki}, {Dermer},
  {Gallo}, {Nakagawa}, {Reynolds}, {Sanders}, {Terashima}, \&
  {Wilson}}]{teng09}
{Teng}, S.~H., {Veilleux}, S., {Anabuki}, N., {Dermer}, C.~D., {Gallo}, L.~C.,
  {Nakagawa}, T., {Reynolds}, C.~S., {Sanders}, D.~B., {Terashima}, Y., \&
  {Wilson}, A.~S. 2009, \apj, 691, 261

\bibitem[{{Teng} {et~al.}(2005){Teng}, {Wilson}, {Veilleux}, {Young},
  {Sanders}, \& {Nagar}}]{teng05}
{Teng}, S.~H., {Wilson}, A.~S., {Veilleux}, S., {Young}, A.~J., {Sanders},
  D.~B., \& {Nagar}, N.~M. 2005, \apj, 633, 664

\bibitem[{{Treister} {et~al.}(2010){Treister}, {Natarajan}, {Sanders}, {Urry},
  {Schawinski}, \& {Kartaltepe}}]{treister10}
{Treister}, E., {Natarajan}, P., {Sanders}, D.~B., {Urry}, C.~M., {Schawinski},
  K., \& {Kartaltepe}, J. 2010, Science, 328, 600

\bibitem[{{Treister} {et~al.}(2009{\natexlab{a}}){Treister}, {Urry}, \&
  {Virani}}]{treister09b}
{Treister}, E., {Urry}, C.~M., \& {Virani}, S. 2009{\natexlab{a}}, \apj, 696,
  110

\bibitem[{{Treister} {et~al.}(2004)}]{treister04}
{Treister}, E. {et~al.} 2004, \apj, 616, 123

\bibitem[{{Treister} {et~al.}(2009{\natexlab{b}})}]{treister09c}
---. 2009{\natexlab{b}}, \apj, 706, 535

\bibitem[{{Tueller} {et~al.}(2008){Tueller}, {Mushotzky}, {Barthelmy},
  {Cannizzo}, {Gehrels}, {Markwardt}, {Skinner}, \& {Winter}}]{tueller08}
{Tueller}, J., {Mushotzky}, R.~F., {Barthelmy}, S., {Cannizzo}, J.~K.,
  {Gehrels}, N., {Markwardt}, C.~B., {Skinner}, G.~K., \& {Winter}, L.~M. 2008,
  \apj, 681, 113

\bibitem[{{Virani} {et~al.}(2006){Virani}, {Treister}, {Urry}, \&
  {Gawiser}}]{virani06}
{Virani}, S.~N., {Treister}, E., {Urry}, C.~M., \& {Gawiser}, E. 2006, \aj,
  131, 2373

\end{thebibliography}
\end{document}